\documentstyle[12pt]{article}
\begin{document}
\title{THE NEW COSMOS}
\author{B.G. Sidharth\\
Centre for Applicable Mathematics \& Computer Sciences\\
B.M. Birla Science Centre, Adarsh Nagar, Hyderabad - 500 063,
India}
\date{}
\maketitle
\begin{abstract}
We review the broad status of cosmology and discuss a model of
fluctuational cosmology in which the universe is created in a
phase transition like phenomenon mimicking inflation, and which
further consistently explains latest observations like the ever
expanding and accelerating feature.
\end{abstract}
\section{Introduction}
The Newtonian Universe was one in which there was an absolute background space
in which the basic building blocks of the universe were situated-these were stars.
This view was a quantum jump from the earlier view, based on the Greek model in which
stars and other celestial objects were attached to transparent material spheres, which
prevented them from falling down.\\
When Einstein proposed his General Theory of Relativity some eighty five years ago, the
accepted picture of the universe was one where all major constituents were stationery.
This had puzzled Einstein, because the gravitational pull of these constituents should
make the universe collapse. So Einstein introduced his famous cosmological constant,
essentially a repulsive force that would counterbalance the attractive gravitational
force. Shortly thereafter there were two dramatic discoveries which completely altered
that picture. The first was due to Astronomer Edwin Hubble, who discovered that the
basic constituents or building blocks of the universe were not stars, but rather
huge conglomerations of stars called galaxies. The second discovery was the fact that these
galaxies are rushing away from each other-far from being static the universe was exploding.
There was no need for the counterbalancing cosmic repulsion any more and Einstein dismissed
this as his greatest blunder.\\
Over the next forty odd years, these observations evolved into the Big Bang theory, according
to which all the matter in the universe possibly some fifteen billion years ago, was
concentrated in a speck, at the birth of the universe, which was characterized by an
inconceivable explosion or bang. This lead to the matter being flung outwards, and that
is what keeps the galaxies rushing outwards even today. In the mid sixties confirmation
for the Big Bang model of the universe came from the detection of a cosmic footprint. The
energy of the initial Big Bang would today still be available in the form of cosmic microwaves,
which accidentally were discovered \cite{r1,r2,r3}.\\
Over the next three decades and more, the Big Bang theory was refined further and further. A
n important question was, would the universe continue to expand for ever, though slowing down,
or would the expansion halt one day and the universe collapse back again. Much depended on
the material content or density of the universe. If there was enough matter, then the
expansion would halt and reverse. If not the universe would expand for ever. However
the observed material content of the universe appeared to be insufficient to halt the
expansion.\\
At the same time there were a few other intriguing observations.
For example the velocities along the radius of a galaxy, instead
of sharply falling off, flattened out. All this led astronomers to
invoke dark matter, that is undetected matter. This matter could
be in the form of black holes within galaxies, or brown dwarf
stars which were too faint to be detected, or even massive
neutrinos which were otherwise thought
to be massless. With dark matter thrown in, it appeared that the universe had sufficient
material content to halt, and even reverse the expansion. That is, the universe would expand
up to a point and then collapse.\\
There still were several subtler problems to be addressed. One was the famous horizon problem.
To put it simply the Big Bang was an uncontrolled or random event and so different parts of
the universe in different directions were disconnected in the earliest stage and so today
need not be the same, just as people in different parts of the world need not wear the
same type of dress. Observation however shows that the universe is by and large uniform,
like people in different countries wearing the same dress-that would not be possible
without some form of intercommunication which would violate Einstein's earlier Special
Theory of Relativity, according to which no signal can travel faster than light. The
next problem was, that according to Einstein, due to the material content in the universe,
space should be curved or roughly speaking bent, whereas the universe appears to be flat.
There were other problems as well. For example astronomers predicted that there should be
monopoles that is, simply put, either only North magnetic poles or only South magnetic
poles, unlike the North South combined magnetic poles we encounter. Such monopoles have
failed to show up.\\
Some of these problems were sought to be explained by what has
been called inflationary cosmology whereby, early on, just after
the Big Bang the explosion was a super fast \cite{r4,r5}.\\
What would happen in this case is, that different parts of the universe, which could not
be accessible by light, would now get connected. At the same time, the super fast expansion
in the initial stages would smoothen out any distortion or curvature effects in space,
leading to a flat universe and in the process also eliminate the monopoles.\\
One other feature that has been studied in detail over the past
few decades is that of structure formation in the universe. To put
it simply, why is the universe not a uniform spread of matter and
radiation? On the contrary it is very lumpy with planets, stars,
galaxies and so on, with a lot of space separating these objects.
This has been explained in terms of fluctuations in density, that
is accidentally more matter being present in a given region.
Gravitation would then draw in even more matter and so on. Such
fluctuations would also cause the cosmic background radiation to
be non uniform or anisotropic. Such fluctuations are in fact being
observed.
\section{Latest Observation}
Everything seemed to have fallen into place. The universe appeared to be well behaved.
However from early 1998 the conventional wisdom of cosmology that had developed over the
past three to four decades, began to be challenged. The work of Perlmutter and co-workers,
as also of Schmidt and co-workers was announced in 1998 and it told a sensational if
different story. They had observed carefully very distant supernovae or exploding stars
and to the disbelief of everyone, it turned out that not only was the universe not slowing
down, but was actually accelerating, which also means that it would keep on expanding
eternally \cite{r6,r7,r8}. This was nothing short of an upheaval, and theorists are
under great pressure to explain all this.\\
Suddenly astronomers were talking about dark energy instead of dark matter. Dark energy is an
unknown and mysterious form of energy that brings into play repulsion, in addition to the
attractive force of gravitation. All this is reminiscent of Einstein's greatest blunder
namely the cosmic repulsion itself.\\
However there is a problem. What is dark energy? Physicists speak of such an energy from
what is called the Quantum Vacuum. The idea here is that there cannot be a background vacuum
with exactly zero energy as exact values of energy are forbidden by Quantum Theory. Only the
average energy could be zero. In other words energy would be fluctuating about a zero value.
This is called a Zero Point Field. As a consequence what happens in the vacuum is that
electrons and positrons are continuously created, out of nothing as it were, but these pairs
are very shortlived. Almost instantaneously they annihilate each other and release energy,
which in turn again manifests itself as electron-positron pairs. These effects could lead
to a cosmic repulsion, but the only problem is that the value of the cosmological constant,
in other words the strength of the cosmic repulsion would be much too high. This would go
against observation. The problem has been known for long as the cosmological constant problem.
So astronomers are also talking about the dark energy, now christened Quintessence, leading
to a mysterious new force of repulsion.\\
Yet another dramatic discovery since 1998 has been made with the
help of the SuperKamiokande experiment in Japan \cite{r9}. This
facility observed solar radiation, in particular for the very
strange, maverick supposedly massless particles, neutrinos. It
turns out  that these particles now possess a miniscule mass,
about a billionth that of an electron. The discovery explains one
puzzle, what has been commonly called the solar neutrino problem.
The point is that we seem to receive much less than the
theoretically expected number of neutrinos from solar radiation.
But the theoretical prediction was made on the basis of the
assumption that neutrinos were massless. Even with the tiniest of
masses, the problem disappears. However these observations
challenge what has come to be known as the Standard Model of
Particle Physics, which takes the masslessness of the neutrino for
granted. Could these massive neutrinos be the elusive dark matter?
The answer is no-- this matter is still much too small to stop the
expansion, which is very well in view of the
latest ever expanding scenario.\\
Another iconoclastic dramatic observation which is gaining
confirmation is that, what is called the fine structure constant,
which scientists have considered to be a sacrosanct constant of
the universe, has infact been slowly decreasing over billions of
years \cite{r10,r11}. Webb and co-workers have confirmed this by
observing the spectrum of light from the distant Quasars and
comparing this with spectra in the vicinity. As the fine structure
constant is made up of the electric charge of the electron, the
speed of light, and the Planck constant, this would mean that one
or some or even all of them are not the sacred constants they have
been taken for, but are slowly changing with time. The
consequences of this are quite dramatic. For instance this would
mean that atoms and molecules in the past were not the same as
their counterparts today - this will be true in future also. This
again would have dramatic implications. According to present
thinking, life as we know it depends on a delicate balance between
the different fundamental constants of nature-otherwise life
itself could neither evolve nor sustain. If the values of these
constants change, so would atoms and molecules and the narrow
limits for life get narrower in time.\\
While attempts are being made to modify the successful inflation
theory and other theories also to try to explain these latest
discoveries there are at least two alternative approaches also
being considered. The first is a theory that has been put forward
in the past few years by Moffat, Albrecht, Magueijo, Barrow and
others that contrary to Einstein's Special Relativity the speed of
light is not a universal constant, but rather it has decreased
over billions of years \cite{r12}.
\section{An Alternative Model}
Another approach was proposed by the author in 1997, in which matter is created at random
from a background Quantum Vacuum or dark energy \cite{r13,r14,r15}. Interestingly the
random nature of creation of the particles take place in a fashion similar to inflation,
though this effect would be pronounced in the earliest stages of the birth of the universe,
as in the inflation model. This model successfully predicted an ever expanding and accelerating
universe. It explains a number of mysterious relations between different physical and
astronomical quantities, for example the radius of the universe, the number of particles
in the universe, the mass and size of a typical elementary particle, the universal
gravitational constant, the speed of light and so on. Many of these puzzling relations
had been known for a long time, but in the absence of an explanation, they had been
dismissed as freak coincidences. In the present model, all these relations follow from
the theory, rather than being accidental. Apart from the fact that this model provides
an explanation for the puzzling time variation of the fine structure constant, it also
gives a mechanism for reconciling the two great irreconcible theories of the twentieth
century, namely Einstein's General Relativity and Quantum Theory. The key to this is the
fact that, in both these theories, space and time were taken to be continuous and smooth,
whereas in this model this is no longer true, though these subtler effects can only be
detected at very very tiny scales.\\
Let us now examine this model in a little more detail \cite{r16}.
Let us first consider the usual Compton scale, $l, \tau$ to be the
lower limit for physics in the sense that within this scale we
encounter unphysical zitterbewegung effects. The Planck scale
itself is a special case of the Compton scale. Remembering now,
that given $N$ particles in the universe, the fluctuation in their
number is $\sim \sqrt{N}$, which takes place within the Compton
scale, we have,
$$\frac{dN}{dt} = \frac{\sqrt{N}}{\tau}$$
On integration we get,
\begin{equation}
T = \frac{\hbar}{mc^2} \sqrt{N}\label{e1}
\end{equation}
which can be easily seen to be valid if $T$ is the age of the
universe.\\
We next invoke the well known equation
\begin{equation}
R \approx \frac{GM}{c^2}\label{e2}
\end{equation}
where $R$ is the radius of the universe and $M$ its mass, given by
\begin{equation}
Nm = M\label{e3}
\end{equation}
where $m$ is the mass of a typical elementary particle, the pion.\\
Differentiating equation (\ref{e2}) we get
\begin{equation}
\frac{dR}{dt} \approx HR\label{e4}
\end{equation}
where $H$ in (\ref{e4}) can be identified with the Hubble
Constant, and using (\ref{e2}) is given by,
\begin{equation}
H = \frac{Gm^3c}{\hbar^2}\label{e5}
\end{equation}
Equation (\ref{e1}), (\ref{e2}) and (\ref{e3}) show that in this
formulation, the correct mass, radius and age of the universe can
be deduced given $N$ as the sole cosmological or large scale
parameter. Equation (\ref{e5}) can be written as
\begin{equation}
m \approx \left(\frac{H\hbar^2}{Gc}\right)^{\frac{1}{3}}\label{e6}
\end{equation}
Equation (\ref{e6}) is quite remarkable: it has been empirically
known as an "accidental" or "mysterious" relation. As observed by
Weinberg, \cite{r17}, this is unexplained; it relates a single
cosmological parameter $H$ to constants from microphysics. In our
formulation, equation (\ref{e6})
is no longer a mysterious coincidence but rather a consequence.\\
If we use (\ref{e5}) and (\ref{e4}) as exact equations we can
conclude that there is no cosmological constant $\Lambda$.
However, as they are not exact equations but rather, order of
magnitude relations, it follows that a small cosmological constant
$\Lambda$ is allowed such that
$$\Lambda \leq 0 (H^2)$$
This is consistent with observation and shows that $\Lambda$ is
very very small - this is the so called cosmological constant
problem \cite{r18}.
But it is explained here.\\
We will not go into further detail, but merely observe that this
is a varying $G$ cosmology which can explain such effects as dark
matter, precession of the perihelion of mercury and so on
(Cf.refs.\cite{r15,r16}for details).\\
We would now like to comment upon the Compton scale and the
fluctuational creation of particles alluded to above. Firstly we
observe that the creation of particles can be deduced within the
context of the Nelsonian Stochastic Theory as the formations of
the analogues of Benard convective cells \cite{r19}. This again is
very much in the spirit of El Naschie's Cantorian space time
\cite{r20,r21,r22,r23,r24}. In this case particles are being
produced out of a background Quantum Vacuum or Zero Point Field
which is a pre space time. First a Nelsonian-Brownian process
alluded to above defines the Planck length while a Brownian random
process with the Planck scale is the fundamental interval leads to
the Compton scale (Cf.ref.\cite{r25} for details).\\
This process is a phase transition, a critical phenomenon. To see
this briefly, let us start with the Landau-Ginsburg equation
\cite{r26}
\begin{equation}
- \frac{\hbar^2}{2m} \nabla^2 \psi + \beta |\psi |^2 \psi = -
\alpha \psi\label{e7}
\end{equation}
Here $\hbar$ and $m$ have the same meaning as in usual Quantum
Theory. It is remarkable that the above equation (\ref{e7}) is
identical with a similar Schrodinger like equation based on
amplitudes, where moreover $|\psi |^2$ is proportional to the mass
(or density) of the particle (Cf. ref.\cite{r16} for details). The
equation in question is,
\begin{equation}
\imath \hbar \frac{\partial \psi}{\partial t} =
\frac{-\hbar^2}{2m'} \frac{\partial^2 \psi}{\partial x^2} + \int
\psi^* (x')\psi (x) \psi (x')U(x')dx',\label{e8}
\end{equation}
In (\ref{e8}), $\psi(x)$ is the probability of a particle being at
the point $x$ and the integral is over a region of the order of
the Compton wavelength. From this point of view, the similarity of
(\ref{e8}) with (\ref{e7}) need not be surprising considering also
that near critical points, due to universality diverse phenomena
like magnetism or fluids share similar mathematical equations.
Equation (\ref{e8}) is shown to lead to the Schrodinger equation
with the particle acquiring a mass (Cf.also ref.\cite{r27}).\\
Infact in the Landau-Ginsburg case the coherence length is given
by
\begin{equation}
\xi = \left(\frac{\gamma}{\alpha}\right)^{\frac{1}{2}} = \frac{h
\nu_F} {\Delta}\label{e9}
\end{equation}
which can be easily shown to reduce to the Compton wavelength (Cf. also ref.\cite{r28}.\\
Thus the emergence of Benard cell like elementary particles from
the Quantum Vacuum mimics the Landau-Ginsburg phase transition. In
this case we have a non local growth of correlations reminiscent
of the standard
inflation theory.\\
The above model apart from mimicking inflation also explains the
so called miraculous large number coincidences for example
(\ref{e1}) or (\ref{e5}) hitherto inexplicable, and further
successfully predicted an ever expanding accelerating universe
amongst other things (Cf.ref.\cite{r29}).

\end{document}